\documentclass[conference,final,10pt,a4paper]{IEEEtran}

\usepackage{cite}
\usepackage[pdftex]{graphicx}
 \graphicspath{{../pdf/}{../jpeg/}}
 \DeclareGraphicsExtensions{.pdf,.jpeg,.png}
\usepackage{pgf}
\usepackage{pgfplots}
\usepackage[cmex10]{amsmath}
\usepackage{algorithmic}
\usepackage{array}
\usepackage{mdwmath}
\usepackage{mdwtab}
\usepackage[font=footnotesize]{subfig}
\usepackage{url}
\usepackage{fixedmultirow}
\usepackage{trfsigns}

\graphicspath{{graphics/}}
\graphicspath{{../Bilder/}}

\usetikzlibrary{calc,trees,positioning,arrows,chains,shapes.geometric,%
 decorations.pathreplacing,decorations.pathmorphing,shapes,%
 matrix,shapes.symbols,plotmarks,decorations.markings,shadows,fit,patterns}
\usetikzlibrary{spy,backgrounds}
\usepgfplotslibrary{groupplots}

\pgfplotsset{compat=1.3}
\tikzset{every picture/.style={>=latex}} % latex arrows
\tikzset{every tick label/.style={font=\footnotesize}} % latex arrows
\tikzset{every axis label/.style={font=\small}} % latex arrows
\pgfsetplotmarksize{2pt}

\tikzset{orientation/.is choice,
    orientation/lr/.style={anchor=west,right=1},
    orientation/lr2/.style={anchor=west,right=2},
    orientation/lrd/.style={anchor=west,below=1},
    orientation/lrd2/.style={anchor=west,below=2},
    orientation/rl/.style={anchor=east,left=1},
    orientation/rl2/.style={anchor=east,left=2},
    orientation/ud/.style={anchor=north,below=1},
    orientation/du/.style={anchor=south,above=1},
    orientation/rld/.style={anchor=east,below=1},
    orientation/rld2/.style={anchor=east,below=2},
}
\tikzstyle{scare} = [
]
\tikzstyle{syslinear} = [
 drop shadow={shadow xshift=.6mm,
              shadow yshift=-.6mm},
 fill=white,
 anchor=west,
 rectangle,
 draw=black,
 minimum height=5mm,
 minimum width=5mm,
 inner xsep=0.5em
]

\tikzstyle{sysnonlinear} = [
 drop shadow={shadow xshift=.8mm,
              shadow yshift=-.8mm},
 fill=white,
 double,
 anchor=west,
 rectangle,
 draw=black,
 minimum height=5mm,
 minimum width=5mm,
 inner xsep=0.5em
]

\tikzstyle{syssource} = [
 anchor=west,
 ellipse,
 draw=black,
 minimum height=1.5ex,
 minimum width=1.5em,
 inner xsep=0.5em
]

\tikzstyle{syssink} = [
 anchor=west,
 ellipse,
 draw=black,
 minimum height=1.5ex,
 minimum width=1.5em,
 inner xsep=0.5em
]

\tikzstyle{syssplit} = [
 fill=black,
 draw=black,
]

\tikzstyle{sysadd} = [
 draw,circle,inner sep=-1pt,
]

\tikzstyle{sysmul} = [
 draw,circle,inner sep=-1pt,
]

\definecolor{MyHSBGreen}{hsb}{0.34065,1,0.91}
\pgfplotscreateplotcyclelist{colors4}{%
 {black!100!white},
 {black!75!white},
 {black!50!white},
 {black!25!white},
}
\pgfplotscreateplotcyclelist{colors5}{%
 {black!100!white},
 {black!75!white},
 {black!50!white},
 {black!25!white},
 {densely dashed, black!100!white},
 {densely dashed, black!75!white},
}
\pgfplotscreateplotcyclelist{colors10}{%
 {black!100!white},
 {black!75!white},
 {black!50!white},
 {black!25!white},
 {densely dashed, black!100!white},
 {densely dashed, black!75!white},
 {densely dashed, black!50!white},
 {densely dashed, black!25!white},
 {densely dotted, black!100!white},
 {densely dotted, black!75!white},
 {densely dotted, black!50!white},
}

\pgfplotscreateplotcyclelist{colorsBERMDDFSE}{%
                 every mark/.append style={fill=black},mark=o\\%
                 every mark/.append style={fill=black},mark=star\\%
                 every mark/.append style={fill=black},mark=triangle\\%
                 every mark/.append style={scale=.7,fill=black},mark=square\\%
                 every mark/.append style={fill=black},mark=diamond\\%
  densely dashed,every mark/.append style={solid},mark=pentagon*\\%
  densely dashed,every mark/.append style={solid},mark=*\\%
  densely dashed,every mark/.append style={solid},mark=triangle*\\%
  densely dashed,every mark/.append style={scale=.7,solid},mark=square*\\%
  densely dashed,every mark/.append style={solid},mark=diamond*\\%
  densely dashed,every mark/.append style={solid},mark=star*\\%
}
\pgfplotscreateplotcyclelist{colors1Empty4Full}{%
                 every mark/.append style={fill=black},mark=o\\%
  densely dashed,every mark/.append style={solid},mark=*\\%
  densely dashed,every mark/.append style={solid},mark=triangle*\\%
  densely dashed,every mark/.append style={scale=.7,solid},mark=square*\\%
  densely dashed,every mark/.append style={solid},mark=diamond*\\%
}

\pgfplotscreateplotcyclelist{colors4Empty4Full}{%
                 every mark/.append style={fill=black},mark=o\\%
                 every mark/.append style={fill=black},mark=triangle\\%
                 every mark/.append style={scale=.7,fill=black},mark=square\\%
                 every mark/.append style={fill=black},mark=diamond\\%
  densely dashed,every mark/.append style={solid},mark=*\\%
  densely dashed,every mark/.append style={solid},mark=triangle*\\%
  densely dashed,every mark/.append style={scale=.7,solid},mark=square*\\%
  densely dashed,every mark/.append style={solid},mark=diamond*\\%
}
\pgfplotscreateplotcyclelist{colors6Empty4Full}{%
                 every mark/.append style={fill=black},mark=o\\%
                 every mark/.append style={fill=black},mark=star\\%
                 every mark/.append style={fill=black},mark=triangle\\%
                 every mark/.append style={scale=.7,fill=black},mark=square\\%
                 every mark/.append style={fill=black},mark=diamond\\%
                 every mark/.append style={fill=black},mark=pentagon\\%
  densely dashed,every mark/.append style={solid},mark=*\\%
  densely dashed,every mark/.append style={solid},mark=triangle*\\%
  densely dashed,every mark/.append style={scale=.7,solid},mark=square*\\%
  densely dashed,every mark/.append style={solid},mark=diamond*\\%
}
\pgfplotscreateplotcyclelist{colors8Empty4Full}{%
                 every mark/.append style={fill=black},mark=o\\%
                 every mark/.append style={fill=black},mark=star\\%
                 every mark/.append style={fill=black},mark=triangle\\%
                 every mark/.append style={scale=.7,fill=black},mark=square\\%
                 every mark/.append style={fill=black},mark=diamond\\%
                 every mark/.append style={fill=black},mark=pentagon\\%
                 every mark/.append style={fill=black},mark=oplus\\%
                 every mark/.append style={fill=black},mark=otimes\\%
  densely dashed,every mark/.append style={solid},mark=*\\%
  densely dashed,every mark/.append style={solid},mark=triangle*\\%
  densely dashed,every mark/.append style={scale=.7,solid},mark=square*\\%
  densely dashed,every mark/.append style={solid},mark=diamond*\\%
}

\pgfplotscreateplotcyclelist{colors}{%
                 every mark/.append style={fill=black},mark=o\\%
                 every mark/.append style={fill=black},mark=star\\%
                 every mark/.append style={fill=black},mark=triangle\\%
                 every mark/.append style={scale=.7,fill=black},mark=square\\%
                 every mark/.append style={fill=black},mark=diamond\\%
                 every mark/.append style={fill=black},mark=pentagon\\%
                 every mark/.append style={fill=black},mark=oplus\\%
                 every mark/.append style={fill=black},mark=otimes\\%
                 every mark/.append style={fill=black},mark=asterisk\\%
                 every mark/.append style={fill=black},mark=+\\%
                 every mark/.append style={fill=black},mark=-\\%
                 every mark/.append style={fill=black},mark=|\\%
                 every mark/.append style={fill=black},mark=x\\%
                 every mark/.append style={fill=black},mark=*\\%
                 every mark/.append style={fill=black},mark=triangle*\\%
                 every mark/.append style={scale=.7,fill=black},mark=square*\\%
                 every mark/.append style={fill=black},mark=diamond*\\%
                 every mark/.append style={fill=black},mark=star*\\%
  densely dashed,every mark/.append style={solid,fill=gray},mark=o\\%
  densely dashed,every mark/.append style={solid,fill=gray},mark=triangle\\%
  densely dashed,every mark/.append style={scale=.7,solid,fill=gray},mark=square\\%
  densely dashed,every mark/.append style={solid,fill=gray},mark=diamond\\%
  densely dashed,every mark/.append style={solid,fill=gray},mark=star\\%
  densely dashed,every mark/.append style={solid,fill=gray},mark=pentagon*\\%
  densely dashed,every mark/.append style={solid,fill=gray},mark=*\\%
  densely dashed,every mark/.append style={solid,fill=gray},mark=triangle*\\%
  densely dashed,every mark/.append style={scale=.7,solid,fill=gray},mark=square*\\%
  densely dashed,every mark/.append style={solid,fill=gray},mark=diamond*\\%
  densely dashed,every mark/.append style={solid,fill=gray},mark=star*\\%
  mark=text,text mark=a\\%
  mark=text,text mark=b\\%
  mark=text,text mark=c\\%
  mark=text,text mark=d\\%
  mark=text,text mark=e\\%
  mark=text,text mark=f\\%
  mark=text,text mark=g\\%
}
\pgfdeclarelayer{background layer}
\pgfdeclarelayer{foreground layer}
\pgfsetlayers{background layer,main,foreground layer}

\newcommand{\ie}{\emph{i.e.}}
\newcommand{\eg}{\emph{e.g.}}
\newcommand{\cf}{\emph{cf.}}

\newcommand{\Eg}{\emph{E.g.}}

\newcommand{\lvec}[1]{\ensuremath{\mathrm{\mathbf{#1}}}}  % latin symbols
\providecommand{\de}[1]{\ensuremath{\mathop{\mathrm{d}}}}

\newcommand{\imag}{\ensuremath{\mathrm{j}}}

\renewcommand{\exp}[1]{\ensuremath{\operatorname{exp}\left\{#1\right\}}}

\newcommand{\SYSlinear}[4]{
 \node[syslinear,at=(#2),orientation=#4] (#1) {#3};
}

\newcommand{\SYSnonlinear}[4]{
 \node[sysnonlinear,at=(#2),orientation=#4] (#1) {#3};
}
\newcommand{\SYSAdd}[5]{
 \node[sysadd,at=(#2),orientation=#5] (#1) {$+$};
 \node[at=(#1),#3=1] (#1Text) {#4};
 \draw[->] (#1Text) -- (#1);
}
\newcommand{\SYSMultiply}[5]{
 \node[sysmul,at=(#2),orientation=#5] (#1) {$\times$};
 \node[at=(#1),#3=1] (#1Text) {#4};
 \draw[->] (#1Text) -- (#1);
}

\newcommand{\SYSPlus}[3]{
 \node[sysadd,at=(#2),orientation=#3] (#1) {$+$};
}

\newcommand{\SYSSplit}[3]{
 \node[coordinate,at=(#2),orientation=#3] (#1) {};
 \draw[syssplit] (#1) circle (1pt);
}

\newcommand{\SYSSampler}[2]{
 \node[coordinate,at=(#2)] (SamplerIn) {};
 \node[coordinate,at=(SamplerIn),right=1.5] (#1) {};
 \node[coordinate,at=(SamplerIn),right=0.5] (SamplerInContact) {};
 \node[coordinate,at=(#1),left=0.5] (#1Contact) {};
 \draw[-o] (SamplerIn) -- (SamplerInContact);
 \draw[o-] (#1Contact) -- (#1);
 \draw (SamplerInContact) -- ++ (30:0.9);
 \draw[thin,<-,>=latex] ($(SamplerInContact)+(0.35,0)$) arc (-20:40:0.5)
 node[above] {$kT$};
}

\newcommand{\SYSLowPass}[3]{
 \node[syslinear,at=(#2),orientation=#3,inner sep=0.2em,thin,] (#1) {
  \begin{tikzpicture}[anchor=center,
                      thick,
                      inner sep=0.1]
   \draw[domain=0:2*pi] plot ({0.9*\x/2/pi}, {(0.4*0.5*0.9*(2+sin(\x r)))});
   \draw[double,thin] ($(0.5,0.4*0.5*0.9*2)-(0.15,0.15)$) -- ++(60:0.4);
   \draw[domain=0:2*pi] plot ({0.9*\x/2/pi}, {(0.4*0.5*0.9*(0+sin(\x r)))});
  \end{tikzpicture}
 };
}

\IEEEoverridecommandlockouts
\title{Nonlinear Trellis Description for Convolutionally Encoded Transmission
Over ISI--channels with Applications for CPM}
\author{
 \IEEEauthorblockN{Fabian Schuh and
                   Johannes B. Huber}%
 \IEEEauthorblockA{Institute for Information Transmission,
                   Friedrich-Alexander-Universit\"at Erlangen--N\"urnberg,
                   Germany\\ 
                   mail: \texttt{\{schuh,\,huber\}@LNT.de}}%
\thanks{This work was supported by Bundesministerium f\"ur Wirtschaft und
        Technologie (BMWi) within the project C-PMSE.}
}

\begin{document}
\maketitle
%%%%%%%%%%%%%%%%%%%%%%%%%%%%%%%%%%%%%%%%%%%%%%%%%%%%%%%%%%%%%%%%%%%%%%%%%%%%%%%
\begin{abstract}
 In this paper we propose a matched decoding scheme for convolutionally encoded
 transmission over intersymbol interference (ISI) channels and devise a
 nonlinear trellis description. As an application we show that for coded
 continuous phase modulation (CPM) using a non--coherent receiver the number of
 states of the super trellis can be significantly reduced by means of a matched 
 non--linear trellis encoder.
\end{abstract}
\begin{IEEEkeywords}
ISI--channel;
continuous phase modulation;
convolutionally encoded transmission;
matched decoding;
non--coherent differential detection;
super trellis
\end{IEEEkeywords}
\IEEEpeerreviewmaketitle
%%%%%%%%%%%%%%%%%%%%%%%%%%%%%%%%%%%%%%%%%%%%%%%%%%%%%%%%%%%%%%%%%%%%%%%%%%%%%%%
\section{Introduction}

Convolutional encoding is an attractive encoding scheme due to its low latency
compared to block encoding. When used for transmission with pulse amplitude
modulation (PAM) schemes over an ISI--channel the receiver has to perform
equalization and decoding either in two separated trellises or jointly in one
super trellis when due to strict latency constraints interleaving and by this
iterative equalization--decoding is prohibited.
We propose a technique to merge the channel encoder with the ISI--channel.
Instead of using ring convolutional codes to integrate the encoder into the
$M$-ary ISI--channel as proposed in~\cite{Rimoldi1995} we describe the ISI
channel and the convolutional encoder by a single non--linear trellis encoder
with binary delay elements. %, only

As an application continuous phase modulation (CPM) schemes are employed. It
poses a class of power--efficient constant--envelope modulation
schemes~\cite{anderson1986digital}. Due to the steady phase transitions the
transmit signal of CPM has good spectral properties.
Non--coherent receivers for CPM are relatively simple and straightforward to
implement when compared to a coherent receiver. Using differential detection
and a matched filter at the receiver one can easily retrieve the transmitted
symbols by means of a decision--feedback equalization (DFE) or
maximum--likelihood sequence estimation (MLSE) using the Viterbi algorithm (VA)
comparable to a PAM transmission over ISI--channels. The main
disadvantage of differential detection is the spectral shaping of the channel
noise. To counteract this performance loss, an additional discrete--time noise
whitening filter is introduced which however further extends the overall
impulse response of the transmission system~\cite{Spinnler1999}.

We introduce coded transmission over ISI--channels using the general concept of
serially concatenated convolutional encoder, mapper, modulator and ISI--channel
in Section~\ref{sec:codedTransmission}. At this point we show that the $m$-ary
channel encoder and the $M$-ary transmission scheme over the ISI--channel can
be combined to a single $m$-ary trellis description representing a non--linear
encoding.
In order to facilitate description we restrict encoder and mapper to $m=2$ and
$M=2^n$; \ie\ each output vector of a rate $\frac{K}{n}$ convolutional encoder
is mapped to one PAM--symbol. Section~\ref{sec:RSSE} shows that a complexity
reduction can be achieved by combining multiple trellis states into
hyperstates~\cite{huber1992trelliscodierung,Spinnler1995}.
We then describe non--coherent reception of CPM signals including differential
detection, matched filtering and noise whitening in Section~\ref{sec:NC-CPM}.
In Section~\ref{sec:simresults} we verify via Monte--Carlo simulations that the
proposed approach gives exactly the same performance with less states compared
to the super trellis of the concatenation of convolutional encoder and the
inherent continuous phase encoding of
CPM~\cite{LIT_Huber_Liu_Globecom87,Rimoldi88}.

%%%%%%%%%%%%%%%%%%%%%%%%%%%%%%%%%%%%%%%%%%%%%%%%%%%%%%%%%%%%%%%%%%%%%%%%%%%%%%%
\section{Convolutionally Encoded Transmission over ISI--channels}\label{sec:codedTransmission}

To introduce matched decoding (MD) we transform the convolutionally encoded
transmission scheme step--by--step. For example consider the serial
concatenation of a rate--$\left(\frac{K}{n}=\frac12\right)$ binary
convolutional encoder, a natural mapper, $M$-ary PAM transmission and an
ISI--channel with $L+1$ channel coefficients $h[k]$ with $k$ denoting the time
index, \cf\ Fig.~\ref{fig:EncMapperNonCohCPM}.
% Here $K$ and $n$ denote the number of input symbols and output symbols, respectively.

%%%%%%%%%%%%%%%%%%%%%%%%%%%%%%%%%%%%%%%%%%%%%%%%%%%%%%%%%%%%%%%%%%%%%%%%%%%%%%%
\subsection{Derivation of the Matched Encoder}\label{sec:codedTransmission-StepByStep}

In the conventional approach one would process the receiver input signal first
by a MLSE or a symbol--by--symbol trellis equalizer for the FIR filter $h[k]$
and forward soft-- or hard--output symbols of this trellis equalization to the
decoder for the channel code. But an optimum receiver however would perform
MLSE over the super trellis decoding the binary channel encoder and the ISI
channel impulse response $h[k]$ of length $L$ jointly. In a straight forward
approach the super trellis would have $Z_\mathrm{enc}\cdot M^L$ states, when
$Z_\mathrm{enc}$ is the number of states of the convolutional encoder.
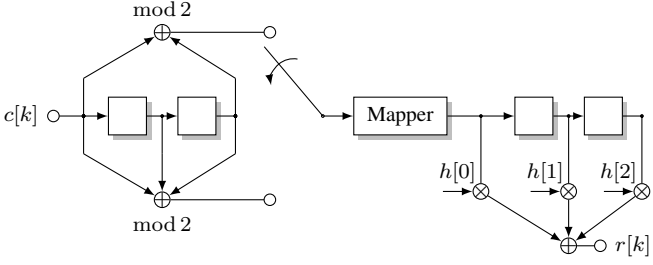
\begin{figure}[ht]\vspace{-2ex}
 \begin{center}
  \begin{tikzpicture}[>=latex,x=1em,y=4ex,font=\footnotesize,inner sep=0.3em,
                      node distance=10mm and 4mm]
   \node at (0,0) (u) {$c[k]$};
   %% Convolutional Encoder
   \node[coordinate,right=of u] (in) {};
   \draw node[syslinear,right=of in] (T1) {};
   \draw node[syslinear,right=of T1] (T2) {};
   \draw node[sysadd, above=of T1.east,xshift=2mm] (g1) {$+$};
   \draw node[sysadd, below=of T1.east,xshift=2mm] (g2) {$+$};
   \node[anchor=south,at=(g1.north)] {$\operatorname{mod}2$};
   \node[anchor=north,at=(g2.south)] {$\operatorname{mod}2$};
   \path (u)       edge[o->] node[coordinate,pos=0.6] (u1) {} (T1)
         (T1)      edge[->]  node[coordinate,midway] (u2) {} (T2)
         (T2.east) edge[-]   node[coordinate,pos=1] (u3) {}  ++(2.5mm,0);
   \draw[fill]  (u1) circle(.5pt) -- ++(0,5mm) edge[->] (g1)
                (u3) circle(.5pt) -- ++(0,5mm) edge[->] (g1)
                (u1) circle(.5pt) -- ++(0,-5mm) edge[->] (g2)
                (u2) circle(.5pt) -- ++(0,-5mm) edge[->] (g2)
                (u3) circle(.5pt) -- ++(0,-5mm) edge[->] (g2);
   \draw[-o] (g1) -- ++(15mm,0);
   \draw[-o] (g2) -- ++(15mm,0);
   \path (T2.east) -- ++(14mm,0) node[coordinate] (encOut) {};
   \path (encOut) edge[-] node[midway,coordinate] (midwayNode) {} ++(130:12mm);
   \draw[<-] (midwayNode) ++(190:3mm) arc (190:80:3mm);
   \node[syslinear,right=of encOut] (Mapper) {Mapper};
   %% FIR Filter
   \draw node[syslinear,at=(Mapper),xshift=15mm] (T3) {};
   \draw node[syslinear,right=of T3] (T4) {};
   \draw (encOut) circle(.5pt) edge[->] (Mapper)
         (Mapper)   edge[->] node[coordinate,midway] (u4) {} (T3)
         (T3)       edge[->] node[coordinate,midway] (u5) {} (T4)
         (T4.east)  edge[-] node[coordinate,pos=1]   (u6) {} ++(2.5mm,0);
   \node[sysmul,at=(u4),yshift=-10mm] (h1) {$\times$};
   \node[sysmul,at=(u5),yshift=-10mm] (h2) {$\times$};
   \node[sysmul,at=(u6),yshift=-10mm] (h3) {$\times$};
   \draw[<-] (h1) -- ++(-5mm,0) node[pos=0.5,anchor=south] {$h[0]$};
   \draw[<-] (h2) -- ++(-5mm,0) node[pos=0.5,anchor=south] {$h[1]$};
   \draw[<-] (h3) -- ++(-5mm,0) node[pos=0.5,anchor=south] {$h[2]$};
    \draw node[sysadd, below=of h2.south,yshift=5mm] (h) {$+$};
   \draw[fill] (u4) circle(.5pt) -- (h1) edge[->] (h)
               (u5) circle(.5pt) -- (h2) edge[->] (h)
               (u6) circle(.5pt) -- (h3) edge[->] (h);
   %% Output
   \draw[-o] (h) -- ++(5mm,0) node[right] {$r[k]$};
  \end{tikzpicture}\vspace{-2ex}
 \end{center}
 \caption{Concatenation of non--coherent CPM with a rate $\frac12$
          convolutional encoder and an ISI--channel.}
 \label{fig:EncMapperNonCohCPM}\vspace{-2ex}
\end{figure}
If the number of output symbols from the encoder can be related to the size of
the modulation alphabet $M$ so that $n=\log_2(M)$ holds, the following
complexity reduction can achieve exactly the same performance by a trellis
description with less states. To see this, note that in each encoding step,
$n-K$ output symbols of the encoder are redundant and depend on $K$ input
symbols. \Eg, in fig.~\ref{fig:EncMapperNonCohCPM} one of the two channel
encoder output symbols contains no further information. We now show, how to
combine the binary channel encoder with the $M$-ary channel impulse response to
form a single binary non--linear trellis encoder.

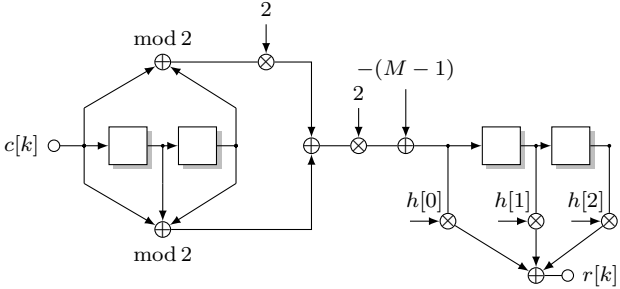
\begin{figure}[ht]\vspace{-2ex}
 \begin{center}
  \begin{tikzpicture}[>=latex,x=1em,y=4ex,font=\footnotesize,inner sep=0.3em,
                      node distance=10mm and 4mm]
    \node at (0,0) (u) {$c[k]$};
    %% Convolutional Encoder
    \node[coordinate,right=of u] (in) {};
    \draw node[syslinear,right=of in] (T1) {};
    \draw node[syslinear,right=of T1] (T2) {};
    \draw node[sysadd, above=of T1.east,xshift=2.0mm] (g1) {$+$};
    \draw node[sysadd, below=of T1.east,xshift=2.0mm] (g2) {$+$};
    \node[anchor=south,at=(g1.north)] {$\operatorname{mod}2$};
    \node[anchor=north,at=(g2.south)] {$\operatorname{mod}2$};
    \path (u)       edge[o->] node[coordinate,pos=0.6] (u1) {} (T1)
          (T1)      edge[->]  node[coordinate,midway] (u2) {} (T2)
          (T2.east) edge[-]   node[coordinate,pos=1] (u3) {}  ++(2.5mm,0);
    \draw[fill]  (u1) circle(.5pt) -- ++(0,5mm) edge[->] (g1)
                 (u3) circle(.5pt) -- ++(0,5mm) edge[->] (g1)
                 (u1) circle(.5pt) -- ++(0,-5mm) edge[->] (g2)
                 (u2) circle(.5pt) -- ++(0,-5mm) edge[->] (g2)
                 (u3) circle(.5pt) -- ++(0,-5mm) edge[->] (g2);
    %% Mod4 addidion
    \node[sysmul,at=(g1.east),xshift=12.5mm] (g1times2) {$\times$};
    \draw[<-] (g1times2) -- ++(0,5mm) node[anchor=south] {$2$};
    \node[sysadd,at=(T2),xshift=15mm] (g1plusg2) {$+$};
    \draw[->] (g1) -- (g1times2) -| (g1plusg2);
    \draw[->] (g2) -| (g1plusg2);
    \node[sysmul,right=of g1plusg2] (bipolar1) {$\times$};
    \draw[<-] (bipolar1) -- ++(0,5mm) node[anchor=south] {$2$};
    \node[sysadd,right=of bipolar1] (bipolar2) {$+$};
    \draw[<-] (bipolar2) -- ++(0,7.5mm) node[anchor=south] {$-(M-1)$};
    %% FIR Filter
    \draw node[syslinear,at=(bipolar2),xshift=10mm] (T3) {};
    \draw node[syslinear,right=of T3] (T4) {};
    \draw[-] (g1plusg2) -- (bipolar1) -- (bipolar2);
    \path (bipolar2) edge[->] node[coordinate,midway] (u4) {} (T3)
          (T3)       edge[->] node[coordinate,midway] (u5) {} (T4)
          (T4.east)  edge[-] node[coordinate,pos=1]   (u6) {} ++(2.5mm,0);
    \node[sysmul,at=(u4),yshift=-10mm] (h1) {$\times$};
    \node[sysmul,at=(u5),yshift=-10mm] (h2) {$\times$};
    \node[sysmul,at=(u6),yshift=-10mm] (h3) {$\times$};
    \draw[<-] (h1) -- ++(-5mm,0) node[pos=0.5,anchor=south] {$h[0]$};
    \draw[<-] (h2) -- ++(-5mm,0) node[pos=0.5,anchor=south] {$h[1]$};
    \draw[<-] (h3) -- ++(-5mm,0) node[pos=0.5,anchor=south] {$h[2]$};
    \draw node[sysadd, below=of h2.south,yshift=5mm] (h) {$+$};
    \draw[fill] (u4) circle(.5pt) -- (h1) edge[->] (h)
                (u5) circle(.5pt) -- (h2) edge[->] (h)
                (u6) circle(.5pt) -- (h3) edge[->] (h);
    %% Output
    \draw[-o] (h) -- ++(5mm,0) node[right] {$r[k]$};
  \end{tikzpicture}\vspace{-2ex}
 \end{center}
 \caption{Equivalent description of the convolutional encoding and ISI--channel
 (\eg\ $M=4$; natural mapping).}
 \label{fig:EncStep2NonCohCPM}\vspace{-2ex}
\end{figure}

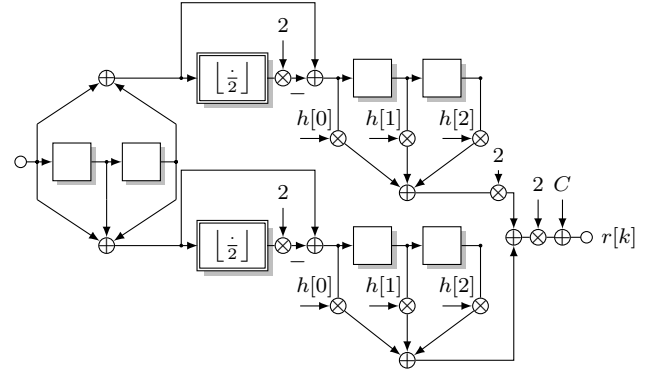
\begin{figure}[ht]\vspace{-1ex}
 \begin{center}
  \begin{tikzpicture}[>=latex,x=10em,y=4ex,font=\footnotesize,inner sep=0.3em,
                      node distance=10mm and 4mm]
    \node at (0,0) (u) {};
    %% Convolutional Encoder
    \node[coordinate,right=of u] (in) {};
    \draw node[syslinear,right=of in,xshift=-3mm] (T1) {};
    \draw node[syslinear,right=of T1] (T2) {};
    \draw node[sysadd, above=of T1.east,xshift=2mm] (g1) {$+$};
    \draw node[sysadd, below=of T1.east,xshift=2mm] (g2) {$+$};
    \path (u)       edge[o->] node[coordinate,pos=0.6] (u1) {} (T1)
          (T1)      edge[->]  node[coordinate,midway] (u2) {} (T2)
          (T2.east) edge[-]   node[coordinate,pos=1] (u3) {}  ++(2.0mm,0);
    \draw[fill]  (u1) circle(.5pt) -- ++(0,5mm) edge[->] (g1)
                 (u3) circle(.5pt) -- ++(0,5mm) edge[->] (g1)
                 (u1) circle(.5pt) -- ++(0,-5mm) edge[->] (g2)
                 (u2) circle(.5pt) -- ++(0,-5mm) edge[->] (g2)
                 (u3) circle(.5pt) -- ++(0,-5mm) edge[->] (g2);
    %% MSB %%%%%%%%%%%%%%%%%%%%%%%%%%%%%%%%%%%%%%%%%%%%%
    \node[sysnonlinear,at=(g1),xshift=12mm] (floorMSB) {$\left\lfloor\frac{\cdot}{2}\right\rfloor$};
    \draw[->] (g1) -- node[coordinate,pos=0.8] (preFloorMSB) {} (floorMSB);
    \node[sysmul,right=of floorMSB,xshift=-3mm] (MSBfloor2) {$\times$};
    \node[sysadd,right=of MSBfloor2,xshift=-2mm] (MSBfloorAdd) {$+$};
    \draw[<-] (MSBfloor2) -- ++(0,5mm) node[anchor=south] {$2$};
    \draw[->] (floorMSB) -- (MSBfloor2) -- (MSBfloorAdd) node[below left] {$-$};
    \draw[->] (preFloorMSB) -- ++(0,10mm) -| (MSBfloorAdd);
    \draw[fill] (preFloorMSB) circle(.5pt);
    %% FIR Filter MSB
    \draw node[syslinear,at=(MSBfloorAdd),xshift=5mm] (T3MSB) {};
    \draw node[syslinear,right=of T3MSB] (T4MSB) {};
    \path (MSBfloorAdd) edge[->] node[coordinate,midway] (u4) {} (T3MSB)
          (T3MSB)       edge[->] node[coordinate,midway] (u5) {} (T4MSB)
          (T4MSB.east)  edge[-] node[coordinate,pos=1]   (u6) {} ++(2.5mm,0);
    \node[sysmul,at=(u4),yshift=-8mm] (MSBh1) {$\times$};
    \node[sysmul,at=(u5),yshift=-8mm] (MSBh2) {$\times$};
    \node[sysmul,at=(u6),yshift=-8mm] (MSBh3) {$\times$};
    \draw[<-] (MSBh1) -- ++(-5mm,0) node[pos=0.5,anchor=south] {$h[0]$};
    \draw[<-] (MSBh2) -- ++(-5mm,0) node[pos=0.5,anchor=south] {$h[1]$};
    \draw[<-] (MSBh3) -- ++(-5mm,0) node[pos=0.5,anchor=south] {$h[2]$};
    \draw node[sysadd, below=of MSBh2.south,yshift=5mm] (MSBh) {$+$};
    \draw[fill] (u4) circle(.5pt) -- (MSBh1) edge[->] (MSBh)
                (u5) circle(.5pt) -- (MSBh2) edge[->] (MSBh)
                (u6) circle(.5pt) -- (MSBh3) edge[->] (MSBh);
    \node[sysmul,at=(MSBh),xshift=12mm] (MSBh2) {$\times$};
    \draw[<-] (MSBh) -- (MSBh2) -- ++(0,3mm) node[anchor=south] {$2$};
    %% LSB %%%%%%%%%%%%%%%%%%%%%%%%%%%%%%%%%%%%%%%%%%%%%
    \node[sysnonlinear,at=(g2),xshift=12mm] (floorLSB) {$\left\lfloor\frac{\cdot}{2}\right\rfloor$};
    \draw[->] (g2) -- node[coordinate,pos=0.8] (preFloorLSB) {} (floorLSB);
    \node[sysmul,right=of floorLSB,xshift=-3mm] (LSBfloor2) {$\times$};
    \node[sysadd,right=of LSBfloor2,xshift=-2mm] (LSBfloorAdd) {$+$};
    \draw[<-] (LSBfloor2) -- ++(0,5mm) node[anchor=south] {$2$};
    \draw[->] (floorLSB) -- (LSBfloor2) -- (LSBfloorAdd) node[below left] {$-$};
    \draw[->] (preFloorLSB) -- ++(0,10mm) -| (LSBfloorAdd);
    \draw[fill] (preFloorLSB) circle(.5pt);
    %% FIR Filter LSB
    \draw node[syslinear,at=(LSBfloorAdd),xshift=5mm] (T3LSB) {};
    \draw node[syslinear,right=of T3LSB] (T4LSB) {};
    \path (LSBfloorAdd) edge[->] node[coordinate,midway] (u4) {} (T3LSB)
          (T3LSB)       edge[->] node[coordinate,midway] (u5) {} (T4LSB)
          (T4LSB.east)  edge[-] node[coordinate,pos=1]   (u6) {} ++(2.5mm,0);
    \node[sysmul,at=(u4),yshift=-8mm] (LSBh1) {$\times$};
    \node[sysmul,at=(u5),yshift=-8mm] (LSBh2) {$\times$};
    \node[sysmul,at=(u6),yshift=-8mm] (LSBh3) {$\times$};
    \draw[<-] (LSBh1) -- ++(-5mm,0) node[pos=0.5,anchor=south] {$h[0]$};
    \draw[<-] (LSBh2) -- ++(-5mm,0) node[pos=0.5,anchor=south] {$h[1]$};
    \draw[<-] (LSBh3) -- ++(-5mm,0) node[pos=0.5,anchor=south] {$h[2]$};
    \draw node[sysadd, below=of LSBh2.south,yshift=5mm] (LSBh) {$+$};
    \draw[fill] (u4) circle(.5pt) -- (LSBh1) edge[->] (LSBh)
                (u5) circle(.5pt) -- (LSBh2) edge[->] (LSBh)
                (u6) circle(.5pt) -- (LSBh3) edge[->] (LSBh);
    %% Output
    \node[sysadd,at=(T2),xshift=49mm,yshift=-10mm] (sumAll) {$+$};
    \draw[->] (LSBh) -| (sumAll);
    \draw[->] (MSBh2) -| (sumAll);
    \node[sysmul,right=of sumAll,xshift=-3mm] (alltimes) {$\times$};
    \node[sysadd,right=of alltimes,xshift=-3mm] (allplus)  {$+$};
    \draw[<-] (alltimes) -- ++(0,5mm) node[anchor=south] {$2$};
    \draw[<-] (allplus) -- ++(0,5mm) node[anchor=south] {$C$};
    \draw[-o] (sumAll) -- (alltimes) -- (allplus) -- ++(4mm,0) node[right] {$r[k]$};
  \end{tikzpicture}\vspace{-2ex}
 \end{center}
 \caption{Replacement of the $\operatorname{mod}2$ addition with the
          non--linear representation using $\operatorname{floor}$ function.}
 %\caption{Equivalent non--linear representation using floor function instead 
 %         of $\operatorname{mod}2$ operation}
 \label{fig:EncStep3NonCohCPM}\vspace{-2ex}
\end{figure}

First, we combine the P/S conversion and the mapper of
Fig.~\ref{fig:EncMapperNonCohCPM}. For clarity, we restrict ourselves to $M=4$,
but note that the concept easily extends to arbitrary $M=2^n$. In this example
$M=4$, \ie\ $n=2$, the upper branch corresponds to the most significant
bit (MSB) whereas the lower branch describes the least significant bit (LSB).
The natural mapping can be applied by multiplying the MSB by $2$ and adding the
LSB. The conversion from unipolar binary symbols $c[k]$ into bipolar symbols
$b[k]$ within an alphabet of size $M$ can be done with $b[k] = (c[k]\cdot2)-1$.
The resulting system is depicted in Fig.~\ref{fig:EncStep2NonCohCPM}.
Recall that the $\operatorname{mod}$ operation can be represented using the
$\operatorname{floor}$ function. In terms of Gaussian notation we can thus
write
\begin{equation}
 x\operatorname{mod} n = x - n\cdot \left\lfloor \frac{x}{n} \right\rfloor.
\end{equation}
In addition we see that the main branch (after the summation of MSB and LSB)
has a multiplication and summation which can be moved behind the convolution.
With $C = -\sum\nolimits_{k=0}^{L}h[k](M-1)$ and the Gauss representation of
the modulo operation we can sketch the transmission system as depicted in
Fig.~\ref{fig:EncStep3NonCohCPM}. Note that now the convolution can be moved
into the MSB branch and LSB branch, respectively, which enables to use binary
delay elements instead of $M$-ary ones.

This representation now has $n$ independent binary branches which all depend on
the same $K$ input values (here $n=2$ and $K=1$). The calculations in each
branch can be combined into a single non--linear filter.

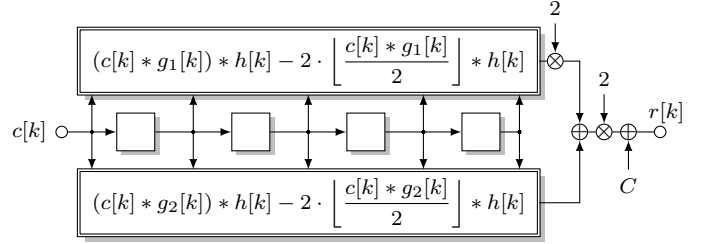
\begin{figure}[ht]\vspace{-2ex}
 \begin{center}
 %\hspace*{-2em}
  \newcommand{\EQMSB}{$\displaystyle
    \left( c[k]*g_1[k] \right)*h[k]
  - 2\cdot\left\lfloor \frac{c[k]*g_1[k]}{2}\right\rfloor * h[k]$}
  \newcommand{\EQLSB}{$\displaystyle
    \left( c[k]*g_2[k] \right)*h[k]
  - 2\cdot\left\lfloor \frac{c[k]*g_2[k]}{2}\right\rfloor * h[k]$}
  \begin{tikzpicture}[>=latex,x=10em,y=4ex,font=\footnotesize,inner sep=0.3em,
                      node distance=10mm and 10mm]
   \node (u) {$c[k]$};
   %% Convolutional Encoder
   \node[coordinate,right=of u,xshift=-12mm] (in) {};
   \draw node[syslinear,right=of in] (T1) {};
   \draw node[syslinear,right=of T1] (T2) {};
   \draw node[syslinear,right=of T2] (T3) {};
   \draw node[syslinear,right=of T3] (T4) {};
   \path (u)       edge[o->] node[coordinate,pos=0.6] (u1) {} (T1)
         (T1)      edge[->]  node[coordinate,midway]   (u2) {} (T2)
         (T2)      edge[->]  node[coordinate,midway]   (u3) {} (T3)
         (T3)      edge[->]  node[coordinate,midway]   (u4) {} (T4)
         (T4.east) edge[-]   node[coordinate,pos=1]    (u5) {}  ++(2.5mm,0);
   \node[sysnonlinear,minimum width=45mm,anchor=south west,at=(T1.west),xshift=-5mm,yshift=5mm]  (MSB) {\EQMSB};
   \node[sysnonlinear,minimum width=45mm,anchor=north west,at=(T1.west),xshift=-5mm,yshift=-5mm] (LSB) {\EQLSB};
   \draw[fill] (u1) circle(.5pt) edge[->] ++(0,5mm)
               (u1) circle(.5pt) edge[->] ++(0,-5mm)
               (u2) circle(.5pt) edge[->] ++(0,5mm)
               (u2) circle(.5pt) edge[->] ++(0,-5mm)
               (u3) circle(.5pt) edge[->] ++(0,5mm)
               (u3) circle(.5pt) edge[->] ++(0,-5mm)
               (u4) circle(.5pt) edge[->] ++(0,5mm)
               (u4) circle(.5pt) edge[->] ++(0,-5mm)
               (u5) circle(.5pt) edge[->] ++(0,5mm)
               (u5) circle(.5pt) edge[->] ++(0,-5mm);
   \node[sysmul,right=of MSB,xshift=-9mm] (mul2MSB) {$\times$};
   \draw[<-] (mul2MSB) -- ++(0,5mm) node[anchor=south] {$2$};
   %% Output
   \node[sysadd,at=(T4),xshift=13mm] (sumAll) {$+$};
   \draw[->] (LSB) -| (sumAll);
   \draw[->] (MSB) -- (mul2MSB) -| (sumAll);
   \node[sysmul,right=of sumAll,xshift=-9mm] (alltimes) {$\times$};
   \node[sysadd,right=of alltimes,xshift=-9mm] (allplus)  {$+$};
   \draw[<-] (alltimes) -- ++(0,5mm) node[anchor=south] {$2$};
   \draw[<-] (allplus)  -- ++(0,-5mm) node[anchor=north] {$C$};%{$\displaystyle-\sum\limits_kh[k](M-1)$};
   \draw[-o] (sumAll) -- (alltimes) -- (allplus) -- ++(5mm,0) node[above] {$r[k]$};
  \end{tikzpicture}\vspace{0ex}
 \end{center}
 \caption{The matched encoder (ME) as a non--linear encoder representation of
          coded non--coherent CPM.}
 \label{fig:EncStep4NonCohCPM}\vspace*{-2ex}
\end{figure}

As an example we use the $4$--state minimum free distance convolutional encoder
with octal representation $\left[ g_{1,\textrm{oct}};\;g_{2,\textrm{oct}}
\right] = \left[ 5_\text{oct};\;7_\text{oct} \right]$ as binary generator
polynomials for the MSB and LSB branch. The resulting, non--linear trellis
encoder is depicted in Fig.~\ref{fig:EncStep4NonCohCPM}.

%%%%%%%%%%%%%%%%%%%%%%%%%%%%%%%%%%%%%%%%%%%%%%%%%%%%%%%%%%%%%%%%%%%%%%%%%%%%%%%
\subsection{Complexity Comparison}\label{sec:codedTransmission-Comparison}

The main advantage of matched encoding is the reduction of the convolution by
the ISI--channel from an $M$-ary input sequence into $\log_2(M)$ binary
parallel convolutions in each branch. As the number of convolutions affect the
calculation of metrics at the receiver but does not influence the number of
resulting MLSE states we will now examine the complexity of trellis
equalization for a matched decoding (MD) receiver and the traditional/serial
super trellis decoding (STD).

%%%%%%%%%%%%%%%%%%%%%%%%%%%%%%%%%%%%%%%%%%%%%%%%%%%%%%%%%%%%%%%%%%%%%%%%%%%%%%%
\subsubsection{Super Trellis Decoding}

In a super trellis we consider channel encoder states and channel states
separately. The channel encoder is defined using generator polynomials with
$\nu$ binary memory elements resulting in $Z_{\text{enc}} = 2^\nu$ states for
the trellis. The channel impulse response can be described using the impulse
response $h[k]$ with $L+1$ filter coefficients. The ISI--channel with $M$-ary
input symbols has $Z_{\text{cha}} = M^L$ states resulting in a total number of
states in the super trellis of
\begin{align}
 Z_{\text{STD}} = Z_{\text{enc}}\cdot Z_{\text{cha}} 
               = 2^\nu \cdot M^L = 2^\nu \cdot 2^{(n\cdot L)}.
\end{align}

%%%%%%%%%%%%%%%%%%%%%%%%%%%%%%%%%%%%%%%%%%%%%%%%%%%%%%%%%%%%%%%%%%%%%%%%%%%%%%%
\subsubsection{Matched Decoding}

There are two differences compared to STD when considering the proposed matched
encoding approach. First, the convolution with the channel impulse response
is done with binary delay elements in contrast to $M$-ary elements. Second,
as the MSB and LSB depend on each other (as of the channel encoder) not all
state transitions are allowed anymore. 
%To see this, consider the matched encoder having only a single possible binary
%input, whereas the super trellis can have a binary input for the encoder and
%an $M$-ary input for the convolution.
%In addition the first filter coefficient $h[0]$ is now incorporated in the
%trellis which increases the number of states by a factor of $2$.
As can be seen from Fig.~\ref{fig:EncStep4NonCohCPM} the total number of delay
elements does not increase although we use binary delay elements, only. Thus,
we still have $2^\nu$ possible states for the binary channel encoder (which is
fully integrated into the non--linear encoder) and $2^L$ possible states for
the convolution resulting in a total number of states of
\begin{align}
 Z_{\text{MD}} &= 2^\nu \cdot 2^{L}.
\end{align}
Recall that for $n=2$ there are two convolutions in parallel for the
computation of the hypothesis.
%%%%%%%%%%%%%%%%%%%%%%%%%%%%%%%%%%%%%%%%%%%%%%%%%%%%%%%%%%%%%%%%%%%%%%%%%%%%%%%
\subsubsection{Comparison}
\begin{table}
 \begin{center}
  \caption{Number of states for non--coherent CPM transmission with $M=4$, $n=2$
           and for the super trellis representation and MD, respectively.}
  \begin{tabular}[ht]{ccccc}\toprule
  Encoder                                                           & $L$ & $Z_{\mathrm{STD}}$ & $Z_{\mathrm{MD}}$ & $G_{\text{MD}}$ \\\midrule
  \multirowbt{5}{*}{$\left[5_\text{oct};\;7_\text{oct}\right]$}     & $0$ & $4$                & $4$               & $1$             \\\cmidrule{2-5}
                                                                    & $1$ & $16$               & $8$               & $2$             \\\cmidrule{2-5}
                                                                    & $2$ & $64$               & $16$              & $4$             \\\cmidrule{2-5}
                                                                    & $3$ & $256$              & $32$              & $8$             \\\cmidrule{2-5}
                                                                    & $4$ & $1024$             & $64$              & $16$            \\%\cmidrule{2-5}
%                                                                    & $5$ & $4096$             & $128$             & $32$            \\
  \midrule
  \multirowbt{5}{*}{$\left[133_\text{oct};\;171_\text{oct}\right]$} & $1$ & $64$               & $64$              & $1$             \\\cmidrule{2-5}
                                                                    & $0$ & $256$              & $128$             & $2$             \\\cmidrule{2-5}
                                                                    & $2$ & $1024$             & $256$             & $4$             \\\cmidrule{2-5}
                                                                    & $3$ & $4096$             & $512$             & $8$             \\\cmidrule{2-5}
                                                                    & $4$ & $16384$            & $1024$            & $16$            \\%\cmidrule{2-5}
%                                                                    & $5$ & $65536$            & $2048$            & $32$            \\
   \bottomrule
   \label{tab:gainStates}\vspace*{-9ex}
  \end{tabular}
 \end{center}
\end{table}

The main advantage of MD compared to STD is the reduction of states
without loss in performance. The resulting trellis still describes the
super trellis but with less states. The gain of this state reduction can be
calculated to
\begin{align}
 G_{\text{MD}} = \frac{Z_{\text{STD}}}{Z_{\text{MD}}}
               = \frac{2^{(n\cdot L)}}{2^{L}}
               = 2^{L(n - 1)}.
\end{align}
Table~\ref{tab:gainStates} summarizes several examples for different encoder
size and channel lengths for the special case of $n=2$ \mbox{($M=4$)}.
Obviously the gain increases with the length of the ISI--channel.

%%%%%%%%%%%%%%%%%%%%%%%%%%%%%%%%%%%%%%%%%%%%%%%%%%%%%%%%%%%%%%%%%%%%%%%%%%%%%%%
\section{Reduced--State Sequence Estimation}\label{sec:RSSE}

We have shown that the super trellis of convolutionally encoded transmission
over ISI--channel can be represented using significantly less states by
parallelizing the $M$-ary convolution. At this point we can use reduced state
sequence estimation (RSSE) to further reduce the number of states at the cost
of small loss in Euclidean distance. In RSSE several MLSE states are combined
into hyperstates~\cite{huber1992trelliscodierung,Spinnler1995} which are then
used for decoding. The partitioning is crucial to the performance of RSSE as it
reduces the Euclidean distance.
We w.o.l.o.g assume that the channel impulse response $h[k]$ is minimum phase
(by application of a proper all--pass filter) allowing use a partitioning
comparable to that of delayed decision--feedback sequence estimation
(DFSE)~\cite{Lee1977,Duel-Hallen1989,Eyuboglu1988,Eyuboglu1989}.

%% optional part
% On an ISI--channel with delay length $L$, DFSE generates the trellis on the
% first $q_h<L$ coefficients only. The successors are cancelled using a decision
% feedback in each state and the state register of the Viterbi algorithm. This
% can be interpreted as a particular solution of RSSE using a DFSE--like
% partitioning.
%% Optional part end
%
In this work we apply this DFSE partitioning to use RSSE for MD of
convolutionally encoded CPM with non--coherent differential detection and noise
whitening as will be described below.

%%%%%%%%%%%%%%%%%%%%%%%%%%%%%%%%%%%%%%%%%%%%%%%%%%%%%%%%%%%%%%%%%%%%%%%%%%%%%%%
\section{Non--Coherent CPM Reception}\label{sec:NC-CPM}

Our application for MD is based on a CPM transmission over an AWGN channel. ISI
results from the differential detection and a noise whitening filter which we
will discuss shortly.
The phase of a CPM transmit signal $s(t)$ is given as
\begin{align}
 \label{eq:cpmphase}\theta( \lvec{a} , t) & = 2\pi h\sum\limits_{k=-\infty}^\infty a[k] \int_{-\infty}^t g(\tilde{t}-kT) d\tilde{t}\\
                                          & = 2\pi h\sum\limits_{k=-\infty}^\infty a[k] q(t-kT) \notag
\end{align}
where $h=\frac{p}{q}$ is the modulation index ($p$ and $q$ relative prime),
$g(t)$ is the frequency pulse of length $L\cdot T$ with the symbol duration $T$
and $\lvec{a} = \left\{ a[k] \right\}$ is a sequence of transmit symbols taken
from the bipolar $M$-ary alphabet $\left\{ \pm1;\;\pm3;\;\dots;\;\pm(M-1)
\right\}$ ($M$ even). An integration of $g(t)$ over time $t$ gives the phase
pulse $q(t)$ with normalization $q(t) = \frac12\; \forall\, t\geq LT$. With an
arbitrary phase--offset $\theta_0$ and the signal energy per modulation
interval $E_s$. The equivalent complex baseband (ECB) transmit signal is given
by
\begin{align}
 \label{eq:cpmecb}s(t)                    & = \sqrt{\frac{E_s}{T}} \exp{\imag\left(\theta\left(\lvec{a},t\right) + \theta_0\right)}\\
                                          & = \sqrt{\frac{E_s}{T}} \exp{\imag2\pi h\sum_{k=-\infty}^\infty a[k] q(t-kT) + \theta_0}.\notag
\end{align}

%%%%%%%%%%%%%%%%%%%%%%%%%%%%%%%%%%%%%%%%%%%%%%%%%%%%%%%%%%%%%%%%%%%%%%%%%%%%%%%
\subsection{Differential Detection}

\begin{figure}[ht]\vspace{-3ex}
 \begin{center}
  \begin{tikzpicture}[x=1em,y=4ex,inner sep=0.5em,font=\footnotesize]
   \node[coordinate] at(0,0) (start) {};
   \node at (start) (st) {$r(t)$};
   \SYSLowPass{ZF}{st.east}{lr}
   \SYSnonlinear{FreqDis}{ZF.east}{Diff.}{lr}
   \SYSMultiply{2pih}{FreqDis.east}{above}{$\frac{1}{2\pi h}$}{lr}
   \SYSlinear{MF}{2pih.east}{$\gamma G^*(f)$}{lr}
   \SYSSampler{Sampler}{MF.east}
   \node[right=1] at (Sampler.east) (akhat) {$\hat{a}[k]$};
   \path 
         (st)         edge[o->] (ZF)
         (ZF)         edge[->] (FreqDis)
         (FreqDis)    edge[->] (2pih)
         (2pih)       edge[->] (MF)
         (Sampler)    edge[-o] (akhat);
   \path ($(FreqDis.south east)-(0,1)$) -- 
          node[midway,coordinate] (diffDeMod) {}
         ($(FreqDis.south west)-(0,1)$);
   \node[draw=gray,rounded corners,fill=black!10!white,anchor=north] at (diffDeMod) (diffmodBlock) {
    \begin{tikzpicture}[font=\footnotesize,inner sep=0.5em,rounded corners=0]
     \node[coordinate] at(0,0) (in) {};
     \SYSnonlinear{arg}{in.east}{$\mathrm{arg}\left\{\cdot\right\}$}{lr}
     \SYSnonlinear{phaseverstg}{arg.east}{phase cont.}{lr}
     \SYSSplit{split}{phaseverstg.east}{lr}
     \node[coordinate,right=1.4] at (split.east) (split2) {};
     \SYSlinear{tau}{split2.south}{$T_d$}{lrd}
     \SYSPlus{ptau}{split2.east}{lr}
     \node[at=(ptau.south east),xshift=3pt] {$-$};
     \node[coordinate,right=1] at(ptau.east) (out) {};
     \path 
         (in) edge[->] (arg)
         (arg) edge[->] (phaseverstg)
         (phaseverstg) edge[-] (split)
         (split) edge[->] (ptau)
         (ptau) edge[-] (out);
     \draw[->] (split) |- (tau);
     \draw[->] (tau) -| (ptau);
    \end{tikzpicture}\vspace{-2ex}
   };
   \draw[decoration=brace,decorate] ($(diffmodBlock.north west)+(0,0.1)$) -- 
                                    ($(diffmodBlock.north east)+(0,0.1)$);
   \draw[double,->,shorten <=1.2ex,shorten >=0.5ex] (diffDeMod) -- (FreqDis.south);
   \node[right,at=(diffmodBlock.east),anchor=west] {
    \begin{tabular}{c}
     \parbox{2cm}{\centering
      results in non--Gaussian non--white noise
      $\Phi_{\mathrm{nn}}(f) \sim f^2$\\\vspace{.5em}
      (for $T_d\rightarrow 0$)
     }
    \end{tabular}
   };
  \end{tikzpicture}\vspace{-2ex}
 \end{center}
 \caption{Non--coherent differential demodulation for CPM.}
 \label{fig:diffDemod}\vspace*{-2ex}
\end{figure}
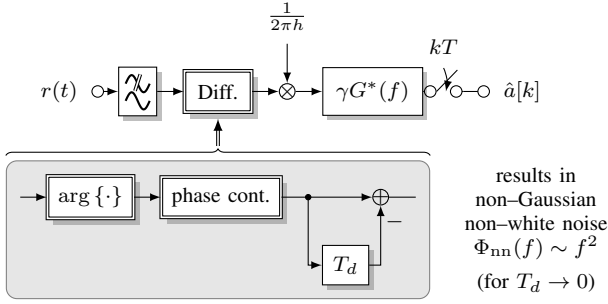

Non--coherent demodulation for additive white Gaussian noise (AWGN) channels
can be implemented using a differential detection as described
in~\cite{Spinnler1999}. In this paper a differential detection for $M$-ary
partial--response ($L>1$) CPM is used. The received signal $r(t)$ is first
band--limited, then the phase of the signal is extracted and unwrapped by a
phase continuation (Phase unwrapping may be implemented by means of sampling at
a sufficiently high frequency, reduction of phase differences
$\operatorname{mod} 2\pi$, and subsequent integration). Finally a
differentiator provides the phase differences between two subsequent samples
separated by $T_d$. The receiver is shown in Fig.~\ref{fig:diffDemod}
(ECB--domain). For an ideal continuous--time differentiation $T_d\rightarrow 0$
holds.
%With the multiplication of $\frac{1}{2\pi h}$ the influence of the modulation
%index is cancelled. 
This demodulator fully inverts the non--linear CPM encoding. Hence a matched
filter $\gamma g^*(-t)$ for transition of continuous to $T$--spaced discrete
time representation may be applied without loss of information on the sequence
of data samples. The parameter $\gamma=\frac{1}{\sqrt{E_g\cdot T}}$ is used for
normalization with the energy of $g(t)$ denoted as $E_g$. Note, that for an
AWGN channel the noise after the matched filter is non--white and non--Gaussian
due to the differential demodulation. This issue is often referred to as
FM--noise (or $f^2$--noise)~\cite{Spinnler1999}.

%%%%%%%%%%%%%%%%%%%%%%%%%%%%%%%%%%%%%%%%%%%%%%%%%%%%%%%%%%%%%%%%%%%%%%%%%%%%%%%
\subsection{Description of Intersymbol Interference}

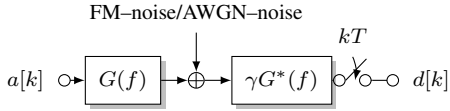
\begin{figure}[ht]\vspace{-2ex}
 \begin{center}
  \begin{tikzpicture}[x=1em,y=4ex,inner sep=0.5em,font=\footnotesize]
   \node (st) {$a[k]$};
   \SYSlinear{gf}{st.east}{$G(f)$}{lr}
   %\SYSAdd{Noise}{gf.east}{above}{$\Phi_\mathrm{nn}(f) \sim f^2 $}{lr}
   \SYSAdd{Noise}{gf.east}{above}{FM--noise/AWGN--noise}{lr}
   \SYSlinear{mf}{Noise.east}{$\gamma G^*(f)$}{lr}
   \SYSSampler{Sampler}{mf.east}
   \node[right=1] at (Sampler.east) (sink) {$d[k]$};
   \path (st) edge[o->] (gf)
         (gf)    edge[->] (Noise)
         (Noise) edge[->] (mf)
         (Sampler)    edge[-o] (sink);
  \end{tikzpicture}\vspace{-1ex}
 \end{center}
 \caption{PAM representation of CPM with differential demodulation (with
          substitute AWGN for derivation of theory).}
 \label{fig:CPMPAMrepres}\vspace*{-2ex}
\end{figure}

As the modulation with differential demodulation only affects the noise power
spectral density we can replace the CPM transmission over an AWGN channel with
an equivalent PAM representation with non--white non--Gaussian noise as
depicted in Fig.~\ref{fig:CPMPAMrepres}.
An optimum receiver for non--coherent CPM would require a continuous--time
whitening filter in front of the matched filter to decorrelate the FM--noise.
As a noise whitening filter would not be stable we neglect the non--white and
non--Gaussian characteristics of the noise at this point and instead develop a
theory for AWGN--channel at first. In~\cite{Spinnler1999} a noise--whitening
filter is introduced after the ``whitened'' matched filter w.r.t AWGN channel
which will be described in detail later on.

In order to achieve a high bandwidth efficiency the common approach for CPM is
to use a Raised Cosine pulse of length $LT$ ($L$-RC pulse) with $L>1$ for
frequency impulse $g(t)$ which results
a) in ISI in the sequence $d[k]$ of $T$-spaced samples of the Matched--Filter
   output (as $G(f)$ does not satisfy the $\sqrt{\text{Nyquist}}$ condition)
   and 
b) in less energy in the spectral side--slopes and thus higher spectral
   efficiency.
Due to ISI equalization has to be performed at the receiver. The ISI can be
described with the energy spectral density $\Phi_{\mathrm{gg}}(f)$ of the
transmit pulse, given by
\begin{align}
 \Phi_{\mathrm{gg}}(f)                  & = \gamma G(f)G^*(f) = \gamma|G(f)|^2 \\
 \intertext{and after $T$--spaced sampling,}
 \Phi_{\mathrm{gg}}[\e^{\imag 2\pi fT}] & \stackrel{\text{def}}{=} \sum\limits_{i=-\infty}^{\infty} \left|\Phi_{\mathrm{gg}}\left(f-\frac{i}{T}\right)\right|^2.
\end{align}
Obviously, $\Phi_{\mathrm{gg}}(f)$ does not fulfill the Nyquist criterion for
$L>1$. For $L$-RC CPM the ISI interacts with $2L-1$ symbols due to
$\Phi_{\mathrm{gg}}[\e^{\imag 2\pi fT}]$.

At this point a Viterbi algorithm can be used to estimate the transmitted
symbols but the noise power--spectral density is still non--white due to the
FM--noise and the matched filter.

%%%%%%%%%%%%%%%%%%%%%%%%%%%%%%%%%%%%%%%%%%%%%%%%%%%%%%%%%%%%%%%%%%%%%%%%%%%%%%%
\subsection{``Whitened Matched Filter'' for AWGN--Channel}

To optimize the performance we now investigate the whitened matched filter
(WMF)~\cite{Forney1972} still assuming AWGN! This additional discrete--time
whitening filter $H_\mathrm{W}(f)$ with an arbitrary phase $\phi(f)$ and the
power spectral density $\Phi_{\mathrm{gg}}[\e^{\imag 2\pi fT}]$ is described by
\begin{align}
 \label{eq:hw}H_\mathrm{W}(f) & = \frac{\e^{\imag\phi(f)}}{\sqrt{\Phi_{\mathrm{gg}}[\e^{\imag 2\pi fT}]}}.
\end{align}
In combination with the receive filter $G^*(f)$ the whitening filter
$H_\mathrm{W}(f)$ results in the whitened matched filter which is a
$\sqrt{\text{Nyquist}}$ function. Using spectral decomposition of the
Z--Transform $\Phi_{\mathrm{gg}}[z] = B[z]\cdot B^*[z^{*-1}]$ we can then
separate the power spectral density $\Phi_{\mathrm{gg}}[\e^{\imag 2\pi fT}]$
into a causal minimum phase $B[z]$ and a non--causal maximum phase part
$B^*(z^{*-1})$.
\begin{align}
 \label{eq:hwmf}
 H_{\mathrm{WMF}}(f) &= \frac{\gamma G^*(f)\e^{\imag\phi(f)}}{\sqrt{\Phi_{\mathrm{gg}}[\e^{\imag 2\pi fT}]}}
                      = \frac{\gamma G^*(f)\e^{\imag\phi(f)}}{B^*[\e^{\imag 2\pi fT}]}
\end{align}
%
% \begin{align}
%  \label{eq:hwmfsqrtNyquist}\sum\limits_{i=-\infty}^{\infty} \left|H_{\mathrm{WMF}}\left(f-\frac{i}{T}\right)\right|^2 = \gamma^2.
% \end{align}
%
% Using this decomposition the overall transfer function can be rewritten as:
% \begin{align}
%  G(f)H_{\mathrm{WMF}}(f) &= G(f)\frac{\gamma G^*(f)}{B^*(\e^{\imag 2\pi fT})} \cdot \frac{B(\e^{\imag 2\pi fT})}{B(\e^{\imag 2\pi fT})}\\
%      &= \gamma \frac{G(f)G^*(f)}{B^*(\e^{\imag 2\pi fT})B(\e^{\imag 2\pi fT})} B(\e^{\imag 2\pi fT})\notag\\
%      &= \gamma \frac{\Phi_{\mathrm{gg}}(f)}{\Phi_{\mathrm{gg}}(\e^{\imag 2\pi fT})} B(\e^{\imag 2\pi fT})\notag
% \end{align}
%
After sampling the overall discrete--time transmission can be summarized using
a single filter $\gamma B[z]$ as depicted in Fig.~\ref{fig:BzFz}. But due to
the differential demodulation the noise is indeed not AWGN but FM--noise
($f^2$--noise).

% \begin{figure}[ht]\vspace{-2ex}
%  \begin{center}
%   \begin{tikzpicture}[x=1em,y=4ex,inner sep=0.5em,font=\footnotesize]
%    \node (st) {$a[k]$};
%    \SYSlinear{B}{st.east}{$\gamma B[z]$}{lr}
%    %\SYSAdd{Noise}{B.east}{above}{$\Phi_\mathrm{nn}(f) \sim f^2 $}{lr}
%    \SYSAdd{Noise}{B.east}{above}{AWGN}{lr}
%    \node[right=1] at (Noise.east) (sink) {$d_W[k]$};
%    \path (st) edge[o->] (B)
%          (B)    edge[->] (Noise)
%          (Noise) edge[->] (sink);
%   \end{tikzpicture}\vspace{-2ex}
%  \end{center}
%  \caption{Equivalent discrete--time representation of a CPM transmission with
%  non--coherent reception using differential detection and whitening filter.}
%  \label{fig:Bz}
% \end{figure}

%%%%%%%%%%%%%%%%%%%%%%%%%%%%%%%%%%%%%%%%%%%%%%%%%%%%%%%%%%%%%%%%%%%%%%%%%%%%%%%
\subsection{Noise Whitening}

In this paper we apply a discrete--time noise whitening filter $F[z]$ after the
``whitened'' matched filter in order to reduce the $f^2$ characteristics of the
noise (see~\cite{Spinnler1999} and Fig.~\ref{fig:BzFz}). Exact noise whitening
would correspond to integration which would cause stability problems. Therefor
we use suboptimum filtering with finite length $L_\mathrm{NW}$.
% In contrast to the optimum receiver input filter
% comprising a continuous--time whitening filter in front of a whitened matched
% filter we here use a $T$--spaced discrete--time whitening filter after the
% ``whitening'' matched filter~\cite{Spinnler1999}.
%
\begin{figure}[ht]\vspace{-2ex}
 \begin{center}
  \begin{tikzpicture}[x=1em,y=4ex,inner sep=0.5em,font=\footnotesize]
   \node (st) {$a[k]$};
   \SYSlinear{B}{st.east}{$\gamma B[z]$}{lr}
   %\SYSAdd{Noise}{B.east}{above}{$\Phi_\mathrm{nn}(f) \sim f^2 $}{lr}
   \SYSAdd{Noise}{B.east}{above}{FM--Noise}{lr}
   \SYSlinear{f}{Noise.east}{$F[z]$}{lr}
   \node[right=1] at (f.east) (sink) {$d_{\textrm{WN}}[k]$};
   \path      (st) edge [o->] (B)
   (B)             edge [->] (Noise)
   (Noise)         edge [->] (f)
   (f)             edge [->] (sink);
  \end{tikzpicture}\vspace{-1ex}
 \end{center}
 \caption{Added $T$--spaced FIR noise whitening filter $F[z]$.}
 \label{fig:BzFz}\vspace*{-2ex}
\end{figure}
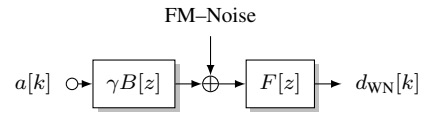
%
% The differentiation in fig~\ref{fig:diffDemod} is causing the colorfulness of
% the noise and has the impulse response
% \begin{equation}
%  h_{\text{diff}}[k] = \delta[k] - \delta[k-1]
% \end{equation}
% whereas $\delta[k]$ denotes the $\delta$--pulse with $\delta[k]=1$ for $k=1$ and
% else $\delta[k]=0$. The autocorrelation then can be calculated to
% \begin{equation}
%  \phi[\kappa] = -\frac12 \delta[k+1] + \delta[k] - \frac12 \delta[k-1].
% \end{equation}
%
\begin{figure}[ht]\vspace{-2ex}
 \begin{center}
  \begin{tikzpicture}[x=1em,y=4ex,inner sep=0.5em,font=\footnotesize]
   \begin{groupplot}[
                group style={group size=2 by 1},
                %group/xticklabels at=edge bottom,
                %group/yticklabels at=edge left,
                group/horizontal sep=1.8em,
                group/vertical sep=0em,
                width=.25\textwidth,
                height=4cm,
                enlargelimits=false,
                grid=both,
                every axis legend/.append style={font={\tiny},nodes={left}},
              		every axis y label/.style={at={(ticklabel cs:0.5)},rotate=90,anchor=center},
                %xlabel near ticks,
               ]
    \nextgroupplot[
                   xmin=-10,xmax=10,
                   ymax=1.2,ymin=-1,
                   xlabel={Index $\kappa$},
                   ytick={-1,0,1},
                  ]
    \pgfplotsset{ylabel={Autocorrelation $\phi[\kappa]$}}
    \addplot+[ycomb] table[x index=0,y index=1] {ACF-noise-L3-h1-4-samps4-RaisedCosineImpulse.csv};
    \nextgroupplot[
                   xlabel={Time index $k$},
                   enlargelimits=false,
                   grid=both,
                   xmin=0,xmax=10,
                   ymax=1.2,ymin=0,
                   every axis legend/.append style={font={\tiny},nodes={left}},
                   axis y line*=right,
                  ]
    \pgfplotsset{ylabel={Filter coefficients $f[k]$}}
    \addplot+[ycomb] coordinates {
                         (0 ,1.0000000e+00)
                         (1 ,1.0692617e+00)
                         (2 ,9.1393741e-01)
                         (3 ,8.2229496e-01)
                         (4 ,7.0743189e-01)
                         (5 ,6.0447927e-01)
                         (6 ,4.9770732e-01)
                         (7 ,3.9663879e-01)
                         (8 ,2.9334665e-01)
                         (9 ,1.9343738e-01)
                         (10,8.8659291e-02)
                     };
   \end{groupplot}
  \end{tikzpicture}\vspace{-2ex}
 \end{center}
 \caption{Auto correlation $\phi[\kappa]$ of the noise and impulse responds of
          the $T$--spaced noise whitening filter $f[k]$ with
          $L_\mathrm{NW}=10$.}
 \label{fig:phikappa}\vspace*{-2ex}
\end{figure}
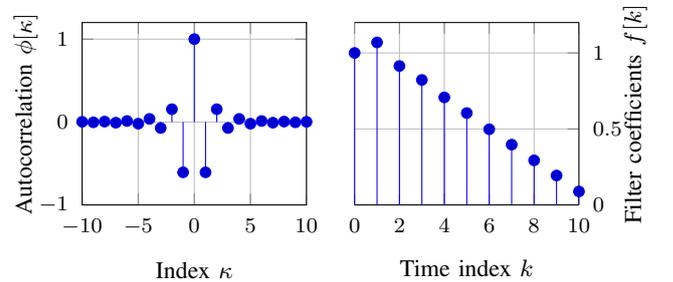
A measurement of the noise characteristics shows that the autocorrelation
$\phi[\kappa]$ depicted in Fig.~\ref{fig:phikappa} (left) clearly shows
correlated noise. We minimize the error variance that results from a prediction
filter with coefficients $p[k]$:
\begin{equation}
 \min\limits_{p[k]} \mathrm{E}\left\{ \left| 
  \phi[\kappa] - \sum\limits_{k=1}^{L_\mathrm{NW}} p[k] \phi[\kappa-k]
 \right|^2 \right\}
\end{equation}
This minimization leads to the Yule--Walker equations with the noise prediction
coefficients $p[k]$:
\begin{equation}
 \sum\limits_{k=1}^{L_{\text{NW}}} p[k] \phi[\kappa-k] = \phi[\kappa] \qquad\forall k=1,2,\dots,L_{\text{NW}}
\end{equation}
% The closed form solution for this linear system then is
% \begin{equation}
%  p[k] = -\frac{L_{\text{NW}}-k+1}{L_{\text{NW}}+1} \qquad\forall k=1,2,\dots,L_{\text{NW}}.
% \end{equation}
The coefficients of the noise whitening filter are then
\begin{equation}
F[z]\;\Ztransf\;f[k] = 
 \begin{cases}
 1     & \text{ for } k = 0;\\
 -p[k] & \text{ for } 1\leq k\leq L_{\text{NW}}.
 \end{cases}
\end{equation}
In the right hand side picture of Fig.~\ref{fig:phikappa} the resulting filter
coefficients for the correlated noise from Fig.~\ref{fig:phikappa} (left) are
depicted.
At the receiver we then have to equalize for ISI $h[k] \ztransf B[z]F[z]$.
Varying the length of the noise whitening filter $L_{\text{NW}}$, we can trade
between the residual colorfulness of the noise and the complexity of the
receiver (in the case of MLSE).

%%%%%%%%%%%%%%%%%%%%%%%%%%%%%%%%%%%%%%%%%%%%%%%%%%%%%%%%%%%%%%%%%%%%%%%%%%%%%%%
\section{Simulation Results}\label{sec:simresults}

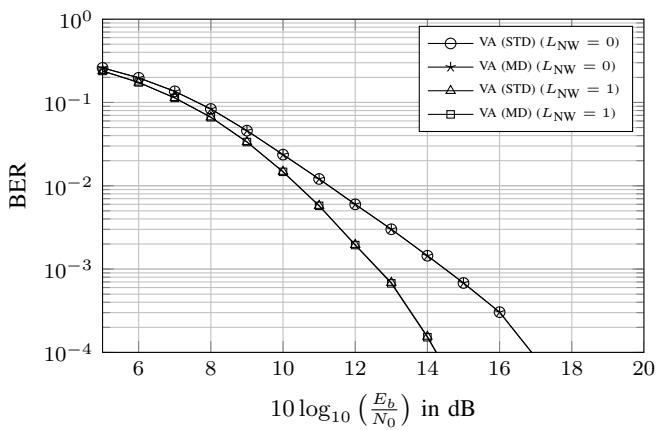
\begin{figure}[ht]\vspace{-2ex}
 \begin{center}
  \begin{tikzpicture}
   \begin{axis}[
                width=.48\textwidth,
                height=6cm,
                xlabel={$10\log_{10}\left(\frac{E_b}{N_0}\right)$ in dB},
                ylabel={BER},
                enlargelimits=false,
                ymode=log,
                cycle list name=colors,
                grid=both,
                xmin=5,xmax=20,
                ymax=1,ymin=1e-4,
                every axis legend/.append style={font={\tiny},nodes={right}},
               ]
    \addplot table[x index=0,y index=5] {MD-vs-ST-Lnw0.data};
    \addplot table[x index=0,y index=6] {MD-vs-ST-Lnw0.data};
    \addplot table[x index=0,y index=5] {MD-vs-ST-Lnw1.data};
    \addplot table[x index=0,y index=6] {MD-vs-ST-Lnw1.data};
    \addlegendentry{VA (STD)             ($L_{\text{NW}}=0$)}
    \addlegendentry{VA (MD)              ($L_{\text{NW}}=0$)}
    \addlegendentry{VA (STD)             ($L_{\text{NW}}=1$)}
    \addlegendentry{VA (MD)              ($L_{\text{NW}}=1$)}
   \end{axis}
  \end{tikzpicture}\vspace{-2ex}
 \end{center}
 \caption{Simulation results for coded non--coherent CPM with the encoder
          polynom $\left[5_\text{oct};\;7_\text{oct}\right]$, $M=4$, 3-RC pulse and $h=\frac14$.}
 \label{fig:resultsMDvsST}\vspace*{-2ex}
\end{figure}

First we will show that the results for MD are indeed exactly the same as for
decoding in the super trellis, \cf, Fig.~\ref{fig:resultsMDvsST}. Here a 4-ary
CPM transmission with modulation index $h=\frac14$ and $L_\mathrm{CPM}=3$ is
used. The lowpass at the receiver limits the signal to the $B_{99.9\%}$
bandwidth of the CPM signal. The ``whitened'' matched filter for AWGN is of
length $20$, whereas the length of the noise whitening filter $F[z]$ is a
parameter. The overall ISI described by $h[k]$ is therefore of length
$L=L_\mathrm{CPM}+L_\mathrm{LNW}-1$. As convolutional encoder, the $\left[
5_\text{oct};\;7_\text{oct} \right]$ code was used.
It becomes clear that the performance of STD is equal to that of MD. The
simulations are conducted only for $L_{\text{NW}} = \left\{ 0;1 \right\}$ due
to fact that STD gets overly complex for longer whitening filters very quickly,
\cf, Table~\ref{tab:gainStates}.

We now investigate the results using MD in combination with RSSE. In
Fig.~\ref{fig:data_md-CPM-3RC-M4-h14-Lnw0-nu7-MD-RSSE-Coded.data} and
Fig.~\ref{fig:data_md-CPM-3RC-M4-h14-Lnw2-nu7-MD-RSSE-Coded.data} the same
coded CPM transmission with non--coherent reception scheme, $L_\textrm{NW}=0$
and $L_\textrm{NW}=2$ but with $64$--state rate--$\frac12$ encoder with
$\left[133_\text{oct};\;171_\text{oct}\right]$ is used.
At the receiver decoding is done using
 a) serial decoding of channel impulse response and channel code 
 b) MLSE decoding using full--state super trellis and
 c) matched decoding with RSSE.
For a) we have two different approaches. The hard decision approach uses a
decision--feedback sequence estimation (DFSE) with $4$ or $16$ states and a
full--state Viterbi algorithm with $2^6$ states to decode the convolutional
encoder. The soft decision approach comprises a symbol--by--symbol detection
using the well--known BCJR algorithm~\cite{BCJR74}, and the soft--input Viterbi
algorithm for channel decoding. One can see that even 4-state MD decoding
supersedes the performance of serial decoding and RSSE converges STD with
increasing number of states.

\begin{figure}[ht]\vspace{-2ex}
 \begin{center}
  \begin{tikzpicture}
   \begin{axis}[
                width=.48\textwidth,
                height=6cm,
                xlabel={$10\log_{10}\left(\frac{E_b}{N_0}\right)$ in dB},
                ylabel={BER},
                enlargelimits=false,
                ymode=log,
                cycle list name=colors,
                grid=both,
                xmin=5,xmax=20,
                ymax=1,ymin=1e-4,
                every axis legend/.append style={font={\tiny},nodes={right}},
               ]
    \addplot table[x index=0,y index=4]         {data_md-CPM-3RC-M4-h14-Lnw0-nu7-MD-RSSE-Coded.data};
    \addplot table[x index=0,y index=5]         {data_md-CPM-3RC-M4-h14-Lnw0-nu7-MD-RSSE-Coded.data};
    \addplot table[x index=0,y index=6]         {data_md-CPM-3RC-M4-h14-Lnw0-nu7-MD-RSSE-Coded.data};
    \addplot table[x index=0,y index=7]         {data_md-CPM-3RC-M4-h14-Lnw0-nu7-MD-RSSE-Coded.data};
    \addplot table[x index=0,y index=8]         {data_md-CPM-3RC-M4-h14-Lnw0-nu7-MD-RSSE-Coded.data};
    \addplot table[x index=0,y index=9]         {data_md-CPM-3RC-M4-h14-Lnw0-nu7-MD-RSSE-Coded.data};
    \addplot table[x index=0,y index=10]        {data_md-CPM-3RC-M4-h14-Lnw0-nu7-MD-RSSE-Coded.data};
    \addplot table[x index=0,y index=11]        {data_md-CPM-3RC-M4-h14-Lnw0-nu7-MD-RSSE-Coded.data};
    \addlegendentry{MD/RSSE (8-states)}
    \addlegendentry{MD/RSSE (16-states)}
    \addlegendentry{MD/RSSE (32-states)}
    \addlegendentry{MD/RSSE (64-states)}
    \addlegendentry{DFSE (04 states)/VA (64 states)}
    \addlegendentry{DFSE (16 states)/VA (64 states)}
    \addlegendentry{BCJR (16 states)/VA (64 states)}
    \addlegendentry{STD (1024 states)}
    %\addplot[dashed] table[x index=0,y index=2] {data_md-CPM-3RC-M4-h14-Lnw0-nu7-MD-RSSE-uncoded.data};
    %\addlegendentry{Uncoded CPM (DFSE 4 states)}
    \node[draw,fill=white,anchor=south west,font={\small}] at (rel axis cs:0.05,0.05) {$L_\mathrm{NW} = 0$};
   \end{axis}
  \end{tikzpicture}\vspace{-2ex}
 \end{center}
 \caption{Simulation results for coded non--coherent CPM with the encoder
          polynom $\left[133_\text{oct};\;171_\text{oct}\right]$, $M=4$, 3-RC pulse,
          $h=\frac14$ and $L_\textrm{NW}=0$.}
 \label{fig:data_md-CPM-3RC-M4-h14-Lnw0-nu7-MD-RSSE-Coded.data}\vspace*{-2ex}
\end{figure}

%%%%%%%%%%%%%%%%%%%%%%%%%%%%%%%%%%%%%%%%%%%%%%%%%%%%%%%%%%%%%%%%%%%%%%%%%%%%%%%
\begin{figure}[ht]\vspace{-2ex}
 \begin{center}
  \begin{tikzpicture}
   \begin{axis}[
                width=.48\textwidth,
                height=6cm,
                xlabel={$10\log_{10}\left(\frac{E_b}{N_0}\right)$ in dB},
                ylabel={BER},
                enlargelimits=false,
                ymode=log,
                cycle list name=colors,
                grid=both,
                xmin=5,xmax=20,
                ymax=1,ymin=1e-4,
                every axis legend/.append style={font={\tiny},nodes={right}},
               ]
    \addplot table[x index=0,y index=4]         {data_md-CPM-3RC-M4-h14-Lnw2-nu7-MD-RSSE-Coded.data};
    \addplot table[x index=0,y index=5]         {data_md-CPM-3RC-M4-h14-Lnw2-nu7-MD-RSSE-Coded.data};
    \addplot table[x index=0,y index=6]         {data_md-CPM-3RC-M4-h14-Lnw2-nu7-MD-RSSE-Coded.data};
    \addplot table[x index=0,y index=7]         {data_md-CPM-3RC-M4-h14-Lnw2-nu7-MD-RSSE-Coded.data};
    \addplot table[x index=0,y index=8]         {data_md-CPM-3RC-M4-h14-Lnw2-nu7-MD-RSSE-Coded.data};
    \addplot table[x index=0,y index=9]         {data_md-CPM-3RC-M4-h14-Lnw2-nu7-MD-RSSE-Coded.data};
    \addplot table[x index=0,y index=1]         {data_md-CPM-3RC-M4-h14-Lnw2-nu7-MD-RSSE-Coded-FullState.data};
    \addlegendentry{MD/RSSE (8-states)}
    \addlegendentry{MD/RSSE (16-states)}
    \addlegendentry{MD/RSSE (32-states)}
    \addlegendentry{MD/RSSE (64-states)}
    \addlegendentry{DFSE (04 states)/VA (64 states)}
    \addlegendentry{DFSE (16 states)/VA (64 states)}
    \addlegendentry{STD (16384-states)}
    %\addplot[dashed] table[x index=0,y index=2] {data_md-CPM-3RC-M4-h14-Lnw2-nu7-MD-RSSE-uncoded.data};
    %\addlegendentry{Uncoded CPM (DFSE 4 states)}
    \node[draw,fill=white,anchor=south west,font={\small}] at (rel axis cs:0.05,0.05) {$L_\mathrm{NW} = 2$};
   \end{axis}
  \end{tikzpicture}\vspace{-2ex}
 \end{center}
 \caption{Simulation results for coded non--coherent CPM with the encoder
          polynom $\left[133_\text{oct};\;171_\text{oct}\right]$, $M=4$, 3-RC pulse, $h=\frac14$ and
          $L_\textrm{NW}=2$.}
 \label{fig:data_md-CPM-3RC-M4-h14-Lnw2-nu7-MD-RSSE-Coded.data}\vspace{-2ex}
\end{figure}

We now investigate the receiver complexity for concatenated equalization and
channel decoding, MD and STD. We compare different channel encodings ISI
channels defined by their number of states. The target bit error rate is
$10^{-3}$ and the receiver complexity is described by the number of states. For
STD and MD with RSSE the receiver complexity is directly given by the number of
states in the super trellis or can be defined, respectively. For concatenated
equalization and decoding the receiver complexity is defined as the sum of
states in the equalization and the decoding, \ie\ $C_\text{serial}=
Z_\mathrm{eq} + Z_\mathrm{enc}$; $C_\text{STD} = Z_\mathrm{eq} \cdot
Z_\mathrm{enc}$ with $Z_\mathrm{eq}$ and $Z_\mathrm{enc}$ is the number of
states in equalization trellis and the channel decoding trellis, represented.
For RSSE the complexity depends on the partitioning so that $C_\text{RSSE} =
2^r$ with arbitrary $r$.

\begin{figure}[ht]%\vspace{-2ex}
 \begin{center}
  \begin{tikzpicture}
   \begin{axis}[
                width=.48\textwidth,
                height=6cm,
                ylabel={receiver complexity $C$},
                xlabel={$10\log_{10}\left(\frac{E_b}{N_0}\right)$ in dB},
                enlargelimits=true,
                ymode=log,
                log basis y=2,
                cycle list name=colors,
                grid=both,
                xmin=12,xmax=21,
                ymin=2,ymax=512,
                every axis legend/.append style={font={\tiny},nodes={right}},
               ]
    %%%%%%%%%%%%%%%%%%%%%%%%%%%%%%%%%%%%%%%%%%%%%%%%%%%%%%%%%%%%%%%%%%%%%%%%%%%
    \addplot coordinates {(1.4545455e+01,32)
                          (1.4545455e+01,16)} 
                          node[pos=0.5,rotate=90,below] {\footnotesize{$L_{\mathrm{NW}}=0$}};
    \addlegendentry{MD 4/16}
    \addplot coordinates {(1.2727273e+01,256)
                          (1.2727273e+01,32)}
                          node[pos=0.5,rotate=90,below] {\footnotesize{$L_{\mathrm{NW}}=1$}}
                          node[pos=0,pin=-100:\footnotesize{STD}] (STDPIN) {}
                          node[pos=1,pin=-100:\footnotesize{MD}] (MDPIN) {}
                          ;
    \addlegendentry{MD 4/64}

    %%%%%%%%%%%%%%%%%%%%%%%%%%%%%%%%%%%%%%%%%%%%%%%%%%%%%%%%%%%%%%%%%%%%%%%%%%%
    % data_md-CPM-3RC-M4-h14-Lnw8-nu3-MD-RSSE-Coded.data
    \addplot coordinates {
                          (1.4242424e+01,4+4)
                          (1.4242424e+01,16+4)
                         };
    \addlegendentry{DFSE/VA 4/$4^{10}$}

    %%%%%%%%%%%%%%%%%%%%%%%%%%%%%%%%%%%%%%%%%%%%%%%%%%%%%%%%%%%%%%%%%%%%%%%%%%%
    % data_md-CPM-3RC-M4-h14-Lnw0-nu3-MD-RSSE-Coded.data
    \addplot coordinates {
                          (1.8030303e+01,4+4)  % DFSE -1/-2
                          (1.7878788e+01,16+4)
                         };
    \addlegendentry{DFSE/VA 4/16}
%     %%%%%%%%%%%%%%%%%%%%%%%%%%%%%%%%%%%%%%%%%%%%%%%%%%%%%%%%%%%%%%%%%%%%%%%%%%%
%     % data_md-CPM-3RC-M4-h14-Lnw2-nu7-MD-RSSE-Coded-BER1e-3.data
%     % data_md-CPM-3RC-M4-h14-Lnw2-nu7-MD-RSSE-Coded-FullState-BER1e-3.data
%     \addplot coordinates {
%                           (1.4848485e+01,4+64) % DFSE-1-VA
%                           (1.4090909e+01,16+64) % DFSE-2-VA
%                          }
%                         %node[pos=0,pin=-10:\footnotesize{STD}] {}
%                         %node[pos=0.5,pin=-10:\footnotesize{$L_{\mathrm{NW}}=1$}] {}
%                         %node[pos=1,pin=-10:\footnotesize{MD}] {}
%                         ;
%     \addlegendentry{DFSE/VA 64/4}

    %%%%%%%%%%%%%%%%%%%%%%%%%%%%%%%%%%%%%%%%%%%%%%%%%%%%%%%%%%%%%%%%%%%%%%%%%%%
    % data_md-CPM-3RC-M4-h14-Lnw0-nu7-MD-RSSE-Coded.data
    \addplot coordinates {
                          (1.7575758e+01,4+64)  % DFSE-1
                          (1.7575758e+01,16+64) % DFSE-2
                         };
    \addlegendentry{DFSE/VA 64/16}

    %%%%%%%%%%%%%%%%%%%%%%%%%%%%%%%%%%%%%%%%%%%%%%%%%%%%%%%%%%%%%%%%%%%%%%%%%%%
    % data_md-CPM-3RC-M4-h14-Lnw0-nu3-MD-RSSE-Coded.data
    \addplot coordinates {
                          (1.7121212e+01,16+4) % BCJR
                         }
                         node[pos=0,pin={-95:\footnotesize{BCJR 4/16}}] {};

    %%%%%%%%%%%%%%%%%%%%%%%%%%%%%%%%%%%%%%%%%%%%%%%%%%%%%%%%%%%%%%%%%%%%%%%%%%%
    % data_md-CPM-3RC-M4-h14-Lnw0-nu7-MD-RSSE-Coded.data
    \addplot coordinates {
                          (1.5909091e+01,16+64) % BCJR
                         }
                         node[pos=0,pin={[pin distance=.2cm]90:\footnotesize{BCJR 64/16}}] {};

    \node[draw,fill=white,anchor=north,font={\small}] at (rel axis cs:0.5,0.95) {$\mathrm{BER} = 10^{-3}$};
   \end{axis}
  \end{tikzpicture}\vspace{-2ex}
 \end{center}
 \caption{Receiver complexity versus SNR for 
          concatenated equalization and decoding, MD and STD.}
 \label{fig:NumbStatesSerial}\vspace{-2ex}
\end{figure}
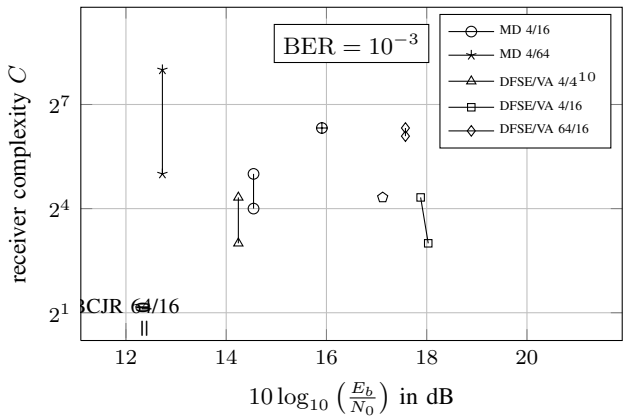

In Fig.~\ref{fig:NumbStatesSerial} we compare MD for a transmission scheme with
$4$ channel encoder states and $16$ or $32$ states in the ISI trellis (see also
Fig.~\ref{fig:resultsMDvsST}). We can see again, that MD performs equally
compared to STD, but with much less states. Additionally the receiver
complexity for concatenated receiver structures are included. The best
performance is achieve with DFSE equalizing a relatively long ISI of length
$10$. The channel encoding and the ISI--channel are described by their number of
states and abbreviated with $Z_\text{enc}/Z_\text{cha}$.

%%%%%%%%%%%%%%%%%%%%%%%%%%%%%%%%%%%%%%%%%%%%%%%%%%%%%%%%%%%%%%%%%%%%%%%%%%%%%%%

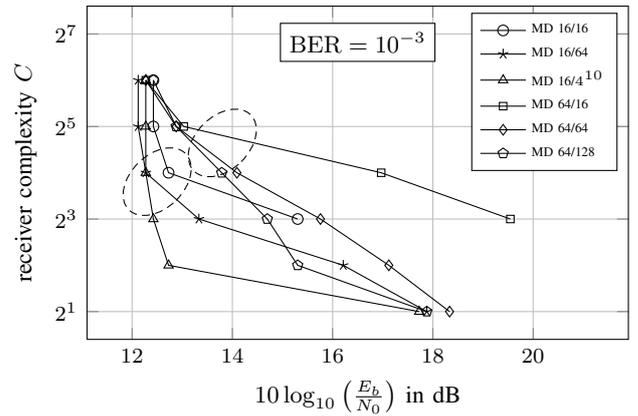
\begin{figure}[ht]\vspace{1ex}
 \begin{center}
  \begin{tikzpicture}
   \begin{axis}[
                width=.48\textwidth,
                height=6cm,
                ylabel={receiver complexity $C$},
                xlabel={$10\log_{10}\left(\frac{E_b}{N_0}\right)$ in dB},
                enlargelimits=true,
                ymode=log,
                log basis y=2,
                cycle list name=colors,
                grid=both,
                xmin=12,xmax=21,
                ymin=2,ymax=128,
                every axis legend/.append style={font={\tiny},nodes={right}},
               ]

    %%%%%%%%%%%%%%%%%%%%%%%%%%%%%%%%%%%%%%%%%%%%%%%%%%%%%%%%%%%%%%%%%%%%%%%%%%%
    % 16 16
    \addplot coordinates {
    (1.5303030e+01, 8)
    (1.2727273e+01,16)
    (1.2424242e+01,32)
    (1.2424242e+01,64)
    };
    \addlegendentry{MD 16/16}
    %%%%%%%%%%%%%%%%%%%%%%%%%%%%%%%%%%%%%%%%%%%%%%%%%%%%%%%%%%%%%%%%%%%%%%%%%%%
    % 16 64
    \addplot coordinates {
    (1.7878788e+01, 2)
    (1.6212121e+01, 4)
    (1.3333333e+01, 8)
    (1.2272727e+01,16)
    (1.2121212e+01,32)
    (1.2121212e+01,64)
    };
    \addlegendentry{MD 16/64}
    %%%%%%%%%%%%%%%%%%%%%%%%%%%%%%%%%%%%%%%%%%%%%%%%%%%%%%%%%%%%%%%%%%%%%%%%%%%
    % data_md-CPM-3RC-M4-h14-Lnw8-nu3-MD-RSSE-Coded.data
    \addplot coordinates {
                          (1.7727273e+01,2)
                          (1.2727273e+01,4)
                          (1.2424242e+01,8)
                          (1.2272727e+01,16)
                          (1.2272727e+01,32)
                          (1.2272727e+01,64)
                         };
    \addlegendentry{MD 16/$4^{10}$}
    %%%%%%%%%%%%%%%%%%%%%%%%%%%%%%%%%%%%%%%%%%%%%%%%%%%%%%%%%%%%%%%%%%%%%%%%%%%
    % 64 16
    \addplot coordinates {
    (1.9545455e+01, 8)
    (1.6969697e+01,16)
    (1.3030303e+01,32)
    (1.2272727e+01,64)
    };
    \addlegendentry{MD 64/16}

    %%%%%%%%%%%%%%%%%%%%%%%%%%%%%%%%%%%%%%%%%%%%%%%%%%%%%%%%%%%%%%%%%%%%%%%%%%%
    % 64 64
    \addplot coordinates {
    (1.8333333e+01, 2)
    (1.7121212e+01, 4)
    (1.5757576e+01, 8)
    (1.4090909e+01,16)
    (1.2878788e+01,32)
    (1.2272727e+01,64)
    };
    \addlegendentry{MD 64/64}

    %%%%%%%%%%%%%%%%%%%%%%%%%%%%%%%%%%%%%%%%%%%%%%%%%%%%%%%%%%%%%%%%%%%%%%%%%%%
    % data_md-CPM-3RC-M4-h14-Lnw2-nu7-MD-RSSE-Coded-BER1e-3.data
    % data_md-CPM-3RC-M4-h14-Lnw2-nu7-MD-RSSE-Coded-FullState-BER1e-3.data
    \addplot coordinates {
                          (1.7878788e+01,2  ) % RSSE - 2-64
                          (1.5303030e+01,4  )
                          (1.4696970e+01,8  )
                          (1.3787879e+01,16 )
                          (1.2878788e+01,32 )
                          (1.2424242e+01,64 )
                          %(1.2121212e+01,1024) % STD
                          }
                          %node[pos=0,pin=-10:\footnotesize{STD}] {}
                          %node[pos=0.5,pin=-10:\footnotesize{$L_{\mathrm{NW}}=1$}] {}
                          %node[pos=1,pin=-10:\footnotesize{MD}] {}
                          ;
    \addlegendentry{MD 64/128}

    \node[draw,fill=white,anchor=north,font={\small}] at (rel axis cs:0.5,0.95) {$\mathrm{BER} = 10^{-3}$};

    \node[coordinate] at (axis cs:12.5,14) (markA) {};
    \draw[densely dashed,rotate around={-45:(markA)}] (markA) ellipse (1em and 1.5em);
    \node[coordinate] at (axis cs:13.8,25) (markB) {};
    \draw[densely dashed,rotate around={-45:(markB)}] (markB) ellipse (1em and 1.5em);
   \end{axis}
  \end{tikzpicture}\vspace{-2ex}
 \end{center}
 \caption{Receiver complexity versus SNR for MD-RSSE.}
 \label{fig:NumbStatesMD}\vspace{-4ex}
\end{figure}

%%%%%%%%%%%%%%%%%%%%%%%%%%%%%%%%%%%%%%%%%%%%%%%%%%%%%%%%%%%%%%%%%%%%%%%%%%%%%%%
As our approach enables the use of RSSE we can compare the performance for
arbitrary receiver complexity for the given target error rate. In
Fig.~\ref{fig:NumbStatesMD} the results for MD and RSSE for a channel encoder
with $16$ or $64$ states and an ISI trellis with $16$ or $64$ states are
compared. It becomes clear that for CPM with non--coherent reception the
performance increases faster when using longer ISI--channel, \ie\ longer noise
whitening filters than with more states in the convolutional encoder. The
figure shows clearly that, when using a $16$ state channel encoder and an
overall channel impulse response of length $10$, only $4$ states at the receiver
are sufficient to supersede a channel encoding with more states and less ISI.

%%%%%%%%%%%%%%%%%%%%%%%%%%%%%%%%%%%%%%%%%%%%%%%%%%%%%%%%%%%%%%%%%%%%%%%%%%%%%%%
\section{Conclusion}\label{sec:conclusion}

In this paper we have shown that it is possible to reduce the number of states
for the super trellis without loss of performance by transforming the $M$-ary
channel convolution into $\log_2(M)$ parallel binary convolutions. Here a
coded non--coherent CPM transmission is used, but as several other
non-interleaved transmission schemes (\ie\ QAM over ISI--channel) can be
represented as a separate channel encoder and a channel impulse response this
approach may be attractive for such %\newpage\noindent
schemes aswell. We
showed that with MD the same performance can be achieved with much less
effort. By using RSSE with DFSE--like partitioning the complexity can be
reduced even further.

The main drawback of the proposed MD approach is that it cannot be combined
with interleaving between channel encoder and modulation, as in convolutional
bit--interleaved coded modulation.

\IEEEtriggeratref{1}
\IEEEtriggercmd{\enlargethispage{1in}}
\bibliographystyle{IEEEtran}
\bibliography{IEEEabrv,main.bib}
\end{document}